\documentclass[reprint,aps,superscriptaddress]{revtex4-2}
\pdfoutput=1
\usepackage{graphicx}
\usepackage{tcolorbox}
\usepackage{bm}
\usepackage{xr}
\usepackage{float}
\usepackage{amsmath}
\usepackage{booktabs}
\usepackage{xcolor}
\usepackage{soul}
\usepackage{cleveref}
\usepackage{svg}
\usepackage[normalem]{ulem}

\usepackage[algo2e]{algorithm2e} 
\setcitestyle{numbers}
\usepackage{geometry}  
\usepackage{adjustbox} 
\usepackage{booktabs}  

\renewcommand{\sout}[1]{}
\SetKwComment{Comment}{/* }{ */}

\begin{document}

\title{Hyperlink communities in higher-order networks}


\author{Quintino Francesco Lotito}
\email{quintino.lotito@unitn.it}
\affiliation{Department of Information Engineering and Computer Science, University of Trento, via Sommarive 9, 38123 Trento, Italy}

\author{Federico Musciotto}
\affiliation{Dipartimento di Fisica e Chimica Emilio Segr\`e, Universit\`a di Palermo, Viale delle Scienze, Ed. 18, I-90128, Palermo, Italy}

\author{Alberto Montresor}
\affiliation{Department of Information Engineering and Computer Science, University of Trento, via Sommarive 9, 38123 Trento, Italy}

\author{Federico Battiston}
\affiliation{Department of Network and Data Science, Central European University, 1100 Vienna, Austria}

\begin{abstract}
\section*{Abstract}
Many networks can be characterised by the presence of communities, which are groups of units that are closely linked. Identifying these communities can be crucial for understanding the system's overall function. Recently, hypergraphs have emerged as a fundamental tool for modelling systems where interactions are not limited to pairs but may involve an arbitrary number of nodes. In this study, we adopt a dual approach to community detection and extend the concept of link communities to hypergraphs. This extension allows us to extract informative clusters of highly related hyperedges. We analyze the dendrograms obtained by applying hierarchical clustering to distance matrices among hyperedges across a variety of real-world data, showing that hyperlink communities naturally highlight the hierarchical and multiscale structure of higher-order networks. Moreover, hyperlink communities enable us to extract overlapping memberships from nodes, overcoming limitations of traditional hard clustering methods. Finally, we introduce higher-order network cartography as a practical tool for categorizing nodes into different structural roles based on their interaction patterns and community participation. This approach aids in identifying different types of individuals in a variety of real-world social systems. Our work contributes to a better understanding of the structural organization of real-world higher-order systems.
\end{abstract}

\maketitle

\section{Introduction}
 From social and technological systems to the human brain, networks have been used to study the structure of a wide range of relational data~\cite{boccaletti2006complex}. 
 Network representations of real-world systems benefit from a variety of methods and algorithms available for their analysis and have proven particularly useful in uncovering complex patterns of connectivity, both at the local and global scales. To investigate the interplay between the structure and dynamics of networks, a key analysis is the identification of their communities~\cite{girvan2002community}. Communities are usually defined as groups of related nodes, with nodes in the same community sharing many more links than nodes in different communities~\cite{fortunato2016community}. The community structure of a system has been linked to its functional sub-units such as protein complexes~\cite{krogan2006global, gavin2006proteome} or different social spheres in social systems~\cite{wasserman1994social, palla2005uncovering, palla2007quantifying}. Many real-world systems are characterised by complex community structure, for instance, hierarchical~\cite{ravasz2002hierarchical, girvan2002community, sales2007extracting, clauset2008hierarchical}, i.e., communities can be recursively split into nested sub-units, and overlapping~\cite{palla2005uncovering, palla2007quantifying}, i.e., nodes might belong to multiple communities at the same time. Yet, traditional community detection algorithms based on hierarchical grouping of nodes often fail to capture the relationships between overlapping groups that characterize real-world networks. 

An alternative approach is that of link communities~\cite{evans2009line, ahn2010link}, which aims to identify groups of interactions rather than nodes. Indeed, nodes can simultaneously belong to multiple groups (for instance, an individual might have a family, a circle of friends, and a team of colleagues). However, interactions typically occur for a singular purpose, such as two individuals connected by a shared interest or a family relation~\cite{ahn2010link}. By looking at the communities of the links of a given node and exploiting a dual network representation where edges are connected if they share a common node~\cite{evans2009line}, it is possible to reconstruct the overlapping modular structure of the interactions of the different nodes. While deviating from the standard definition of communities, which requires them to be groups of nodes instead of edges, link communities are able to explain the seemingly opposing structural principles of overlapping communities and hierarchy.

While widely used, traditional graph representations are not always able to capture accurately the patterns of interactions that occur in the real world~\cite{lambiotte2019networks}. Indeed, networks can only properly represent dyadic relationships, where pairs of nodes are connected by links~\cite{battiston2020networks, torres2021why, battiston2022higher}. In recent years, however, systems such as cellular networks~\cite{klamt2009hypergraphs}, brain networks~\cite{petri2014homological,giusti2016two}, social systems~\cite{benson2016higher}, ecosystems~\cite{grilli2017higher}, human face-to-face interactions~\cite{cencetti2021temporal}, and collaboration networks~\cite{patania2017shape} have all demonstrated that a significant portion of interactions involves three or more nodes at the same time. Mathematical frameworks like hypergraphs~\cite{berge1973graphs} are well-suited to explicitly encode such higher-order systems, where hyperedges of various sizes can represent structured relationships among multiple units typically known as higher-order interactions~\cite{battiston2020networks, torres2021why, battiston2022higher}. Higher-order network approaches have been applied to a range of systems and proven effective to understand their behaviour, including the emergence of new collective phenomena~\cite{battiston2021physics} in diffusive~\cite{schaub2020random,carletti2020random}, synchronization~\cite{skardal2020higher,millan2020explosive, lucas2020multiorder,gambuzza2021stability, zhang2022higher}, spreading~\cite{iacopini2019simplicial,chowdhary2021simplicial,neuhauser2020multibody}  and evolutionary~\cite{alvarez2021evolutionary} processes. To investigate the higher-order organization of real-world systems, different static~\cite{courtney2016generalized,young2017construction,chodrow2020configuration, ruggeri2022principled} and growing~\cite{kovalenko2021growing,millan2021local} models have been proposed. Moreover, several network measures and concepts have been adapted to consider the presence of higher-order interactions, such as higher-order clustering~\cite{benson2018simplicial},  centrality~\cite{benson2019three, tudisco2021node}, motifs~\cite{lotito2022higher, lotito2023exact},
spectral measures~\cite{krishnagopal2021spectral},
tools for network backboning~\cite{musciotto2021detecting, musciotto2022identifying} and reconstruction~\cite{young2021hypergraph}, providing insights into the structure and function of higher-order systems. 

While there have been some recent efforts to understand the mesoscale organization of real-world hypergraphs~\cite{wolf2016advantages,vazquez2009finding,carletti2021random,eriksson2021choosing, chodrow2021generative,chodrow2022nonbacktracking, contisciani2022inference,ruggeri2023generalized}, this topic remains relatively unexplored. In particular, link communities have been shown to provide new insights into the mesoscale structure of real-world pairwise networks, yet the modular organization of interactions in higher-order systems has not been investigated. Here, we propose a framework to detect hyperlink communities, which allows the identification of clusters of closely related higher-order interactions of arbitrary size. We illustrate the applicability of our method on a variety of real-world higher-order systems, unravelling distinct organization principles revealed at different topological scales. We characterize the patterns of connections of system units across hyperedges belonging to different clusters, revealing a complex overlapping modular organization which can be conveniently described through hypergraph cartography. Our work contributes to a better understanding of the real-world organization of higher-order systems.  

\section{Hyperlink communities}
\begin{figure}
    \centering
    \includegraphics[width=\linewidth]{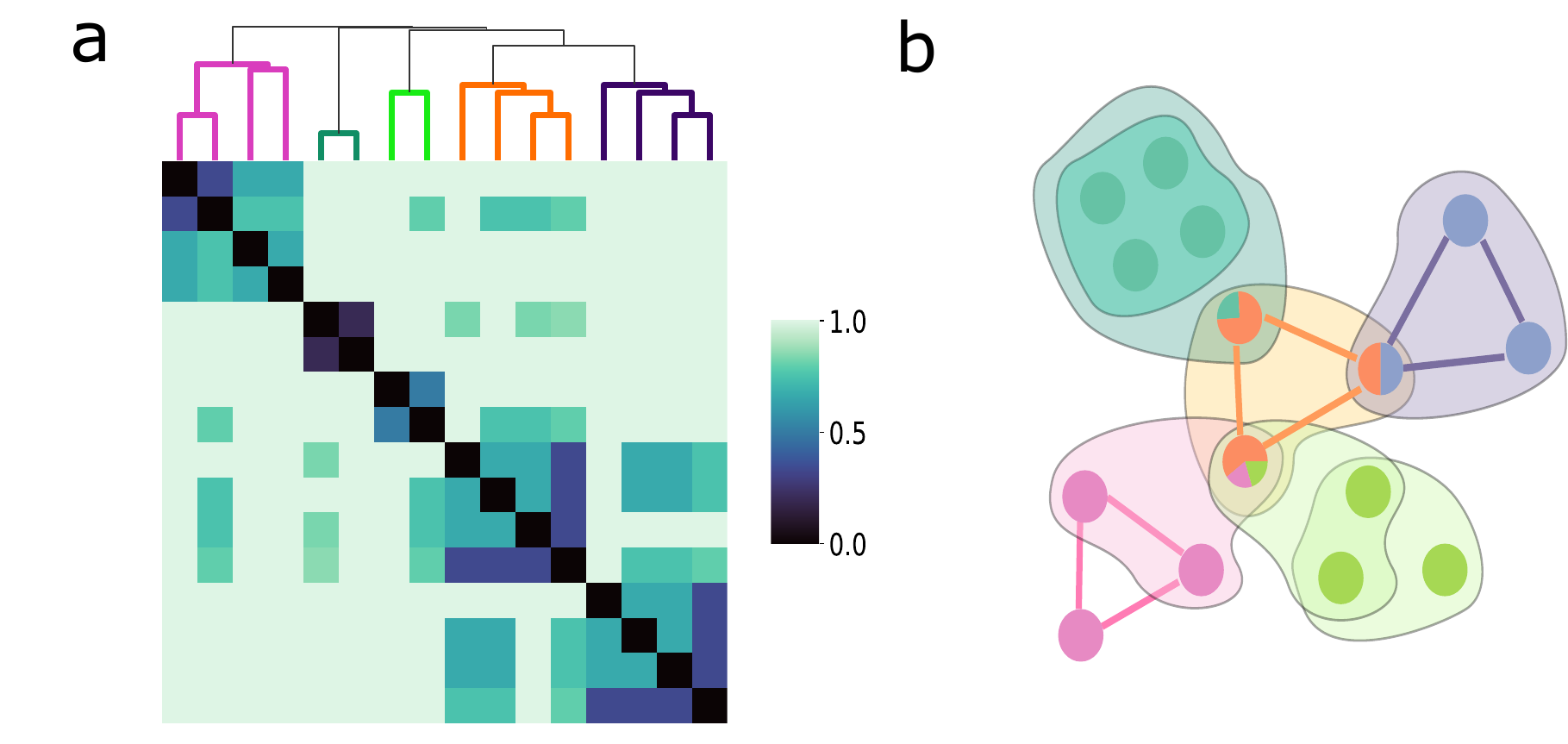}
    \caption{\textbf{Hyperlink communities and their properties.} Hyperlink communities group interactions to describe the mesoscale structure of a hypergraph. This approach is able to explain both the hierarchical organization of hyperedges and the overlap of communities among nodes.
    \textbf{a)} We perform hierarchical clustering on the hyperlinks of an observed hypergraph, considering their Jaccard distance. The output of such clustering is a dendrogram in which the leaves are the hyperlinks and the branches are the hyperlink communities. The dendrogram can be cut at different thresholds, each threshold potentially giving a meaningful community structure as output. \textbf{b)} After the cut, each hyperlink is uniquely assigned to a specific community. Nodes are then assigned to the set of communities to which the hyperlinks in which they are active belong. As a result, a single node may belong to multiple communities simultaneously.}
    \label{fig:fig1}
\end{figure}

The notion of hyperlink communities extends to hypergraphs the idea of describing the mesoscale structure of a system
by grouping (higher-order) interactions instead of nodes. While atypical, this approach can describe the hierarchical organization of hyperedges (Fig.~1a) and node community overlap (Fig.~1b) as two aspects of the same phenomenon. In a way similar to the seminal work on link communities~\cite{ahn2010link}, we define a distance measure between hyperlinks and perform hierarchical clustering on top of them to obtain hyperlink communities (Fig.~1a). Being each hyperlink encoded as a set of interacting nodes, a natural way to compute the distance between two hyperlinks $A$ and $B$ is to consider their Jaccard distance, defined as $J(A, B) = 1 - \frac{|A \cap B|}{|A \cup B|}$. More information about the choice of this distance measure can be found in Appendix~\ref{supp:distance}. To optimize the computation of distances, we precompute and cache sets of hyperedges with at least one common node. This strategy allows our distance algorithm to execute the more time-consuming steps of set-union and set-intersection only when strictly necessary. It is also worth noting that this step can be easily parallelized or distributed in the case of very large hypergraphs. We perform single-linkage hierarchical clustering on top of the distance matrix of the hyperlinks. The clustering procedure follows a bottom-up approach: it starts by assigning each hyperlink to its own cluster and then merges the clusters of hyperlinks with the smallest average distance until all the hyperlinks are part of a single cluster. The output of hierarchical clustering applied on such hyperlinks is a dendrogram. The dendrogram is a graphical representation of the hierarchical structure of the communities formed during the clustering process. In such a dendrogram, leaves are hyperlinks from the observed hypergraph and branches are hyperlink communities. Moreover, the height in the dendrogram of each branch provides additional information about the strength of the merged communities. For a formal description of the process of building the dendrogram, we refer to algorithm~\ref{alg:get_dendrogram}. The dendrogram can be cut at different heights, or thresholds, giving as output different meaningful community structures, typically revealing distinct multiscale organization of hyperlink communities. 

\begin{algorithm2e}
\caption{Hierarchical clustering on hyperlinks}
\label{alg:get_dendrogram}
\KwData{a hypergraph $\mathcal{H} = (V,E)$}
\KwResult{the dendrogram of hyperlink communities}
\texttt{\\}
$\mathcal{D} \gets$ matrix of pairwise distances of hyperlinks;

clusters $\gets$ each hyperlink is assigned to a singleton cluster
\Comment*[r]{The leaves of the dendrogram.}

\While{number of clusters $>1$}{
  merge the clusters with the smallest distance
  \Comment*[r]{The branches of the dendrogram. The height in the dendrogram of a branch is the distance between the two merged clusters.}
}
\end{algorithm2e}

While after a cut each hyperlink is uniquely assigned to a community, nodes inherit the community memberships of all the interactions they participate in. Nodes might participate simultaneously in several communities, although with different strengths. Therefore, another property of hyperlink communities is the ability to naturally extract overlapping communities of the nodes of a hypergraph (Fig.~1b).

Explicitly treating link communities at a higher-order level has several advantages with respect to analyzing lower-order projections of hypergraph datasets. For instance, if we slightly modify the hypergraph in Fig.~1b by adding a new large hyperedge (Fig.~2), its clique-projection is heavily affected, since a large clique is introduced in the system. From a modelling perspective, adding this large clique destroys any hierarchical architecture, while from a computational perspective, it adds a huge number of new edges to consider in the computation. Our higher-order approach, by contrast, does not suffer from such drawbacks.

\begin{figure}
    \centering
\includegraphics[width=\columnwidth]{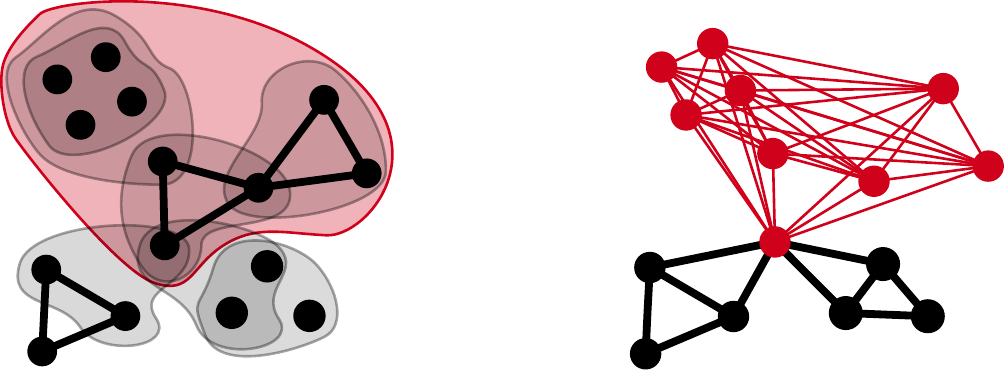}
    \caption{\textbf{Drawbacks of lower-order approaches to higher-order data.} If a large hyperedge (the red one) is added to a hypergraph, it can significantly affect its clique-projection, destroying the ability of low-order community detection tools to capture any hierarchical architecture and introducing a large number of new edges to consider in computation. These drawbacks are resolved when higher-order data are handled using a higher-order approach.}
    \label{fig:fig2}
\end{figure}

\section{Results}
In order to study real-world hypergraphs, we gathered a collection of freely available datasets of systems with group interactions. The datasets come from a variety of domains, including face-to-face proximity contacts in a primary school, in a high school, in a hospital and in a group of baboons ~\cite{stehle2011high, vanhems2013estimating, mastrandrea2015contact, benson2018simplicial,gelardi2020measuring}, e-mail exchange (Enron)~\cite{benson2018simplicial}, biology (NDC classes, i.e., class labels applied to drugs)~\cite{benson2018simplicial} and co-authorship in the physics area (PACS3: Atomic and Molecular Physics, Arxiv physics and society)~\cite{pacs_data}. More information about the datasets used in our experiments can be found in Appendix~\ref{supp:data}. The code of the experiments is freely available~\cite{software2023hyperlink} and it is implemented as part of hypergraphx~\cite{lotito2023hypergraphx}, an open source python library for higher-order network analysis.

\subsection{Multiscale properties of hyperlink communities}
As previously mentioned, the output of the hierarchical clustering algorithm applied to the distance matrix of the hyperlinks is a dendrogram in which the leaves are the hyperlinks, the branches are hyperlink communities, and the height of a branch encodes information about the strength of a hyperlink community. By cutting the dendrogram at different thresholds, we can obtain different community structures, each of which may provide valuable insights into the organization of the hypergraph. This feature is useful for performing analyses of a system at different scales. For example, in a time-varying system, a student might interact more with students sitting nearby during a lecture, with other students in the same class during a break, and with students from other classes during lunch. When aggregated over time, such a variety of patterns might lead to a complex multiscale organization of interactions. The shape of the hierarchical clustering dendrogram depends on the patterns of overlap between hyperlinks. To illustrate how hypergraphs can have very different hierarchies of hyperlinks, depending on the distribution of their overlap distances, in Fig.~3 we show two examples of dendrograms (and their corresponding
distance matrix) of hyperlink communities from real-world hypergraphs: one representing proximity group interactions among baboons and the other representing drugs and their associated class labels. In the following, we show that these two examples are indeed representative of two broader classes of real-world hypergraphs.

\begin{figure*}
    \centering
    \includegraphics[width=\linewidth]{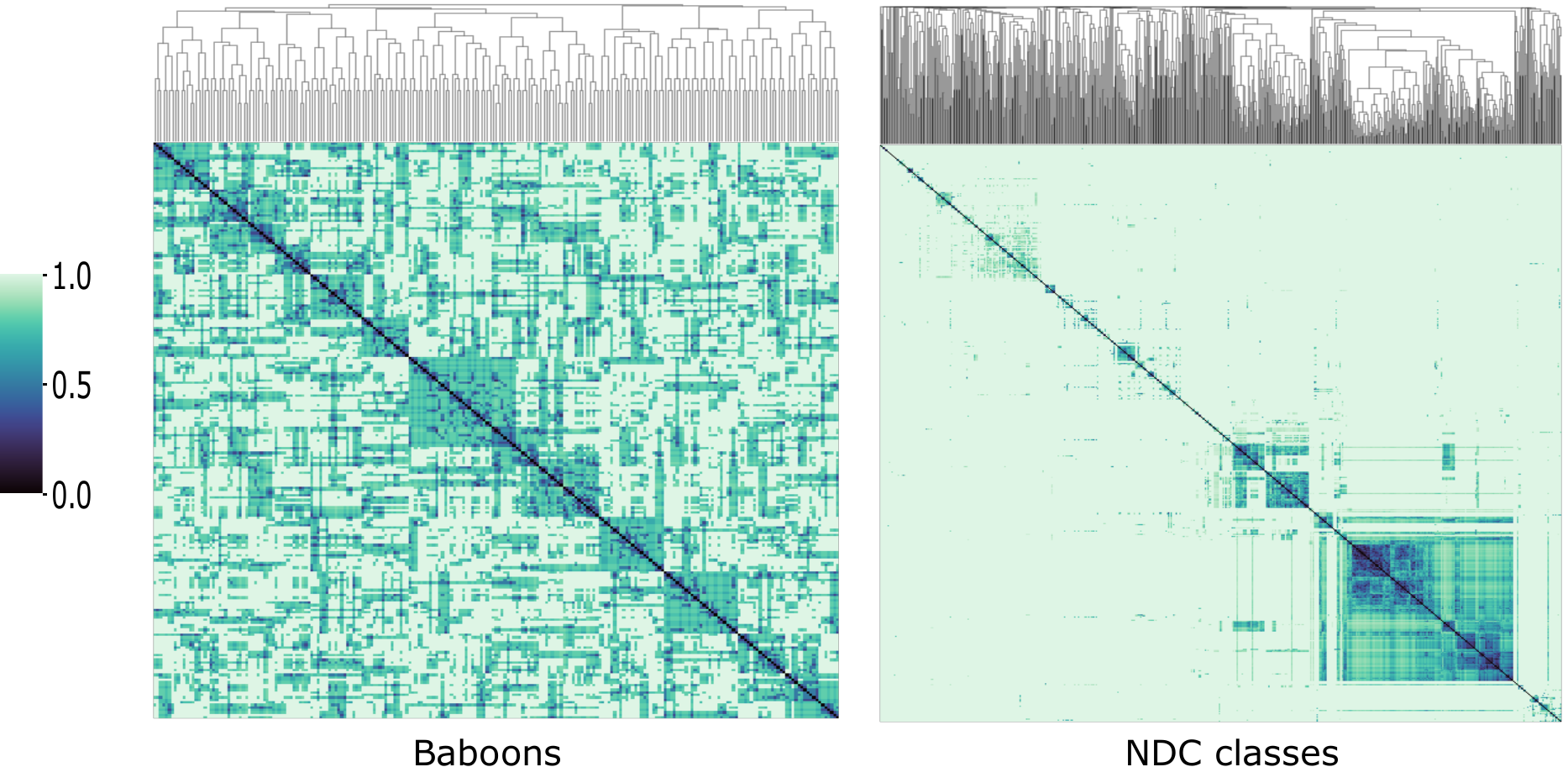}
    \caption{\textbf{Hierarchical clustering of hyperlinks in real-world hypergraphs.} We provide two examples of dendrograms (and their corresponding
distance matrix) of hyperlink communities from real-world hypergraphs: one representing proximity group interactions among baboons, and the other representing affiliations between drugs and class labels applied to each drug. Hypergraphs can show very different hierarchies of hyperlinks, due to different statistics of their overlap distances. In particular, we identified two broader classes of real-world hypergraphs, of which these two examples are representative.}
    \label{fig:fig3}
\end{figure*}

To describe and build a profile of the hierarchy of real-world hypergraphs, we can measure the number of hyperlink communities at different increasing thresholds. The specific scaling of the number of hyperlink communities can be interpreted as a fingerprint of the hierarchy of group interactions in real-world systems. In Fig.~4a, we show such profiles for several datasets from different domains, highlighting the emergence of different behaviours. Data on social proximity, such as contacts within hospitals and schools, reveals distinct spikes at certain thresholds (red lines). These spikes correlate with a prevalent recurrence of specific overlapping patterns among hyperlinks. This behaviour indicates that the dendrograms are hierarchically organized in a few, important distinct topological levels. Instead, affiliation and co-authorship data show a smoother decreasing curve (blue lines). While such datasets approach $0$ at differing rates, they do not exhibit significant spikes. This behaviour indicates that smoother hierarchical transitions are associated with the different system scales. Our multiscale analysis highlights the emergence of two families of real-world hypergraphs, distinguished by their scaling profiles. This characterization of hypergraphs is in agreement with purely local methods, such as motif analysis~\cite{lotito2022higher}.

Statistics about the hyperlink communities obtained by cutting the dendrogram at different thresholds can vary significantly. Fig.~4b shows the distributions of the size of hyperlink communities at three different significant cuts for a subset of the selected datasets, with examples from the two identified classes of real-world hypergraphs. The community structure can change significantly across different scales. This demonstrates that not only do the dendrograms encode information about a hypergraph at multiple scales, but it is also valuable to analyze each level since they can provide distinct insights into the global organization of a system. 

\begin{figure*}
    \centering
    \includegraphics[width=\linewidth]{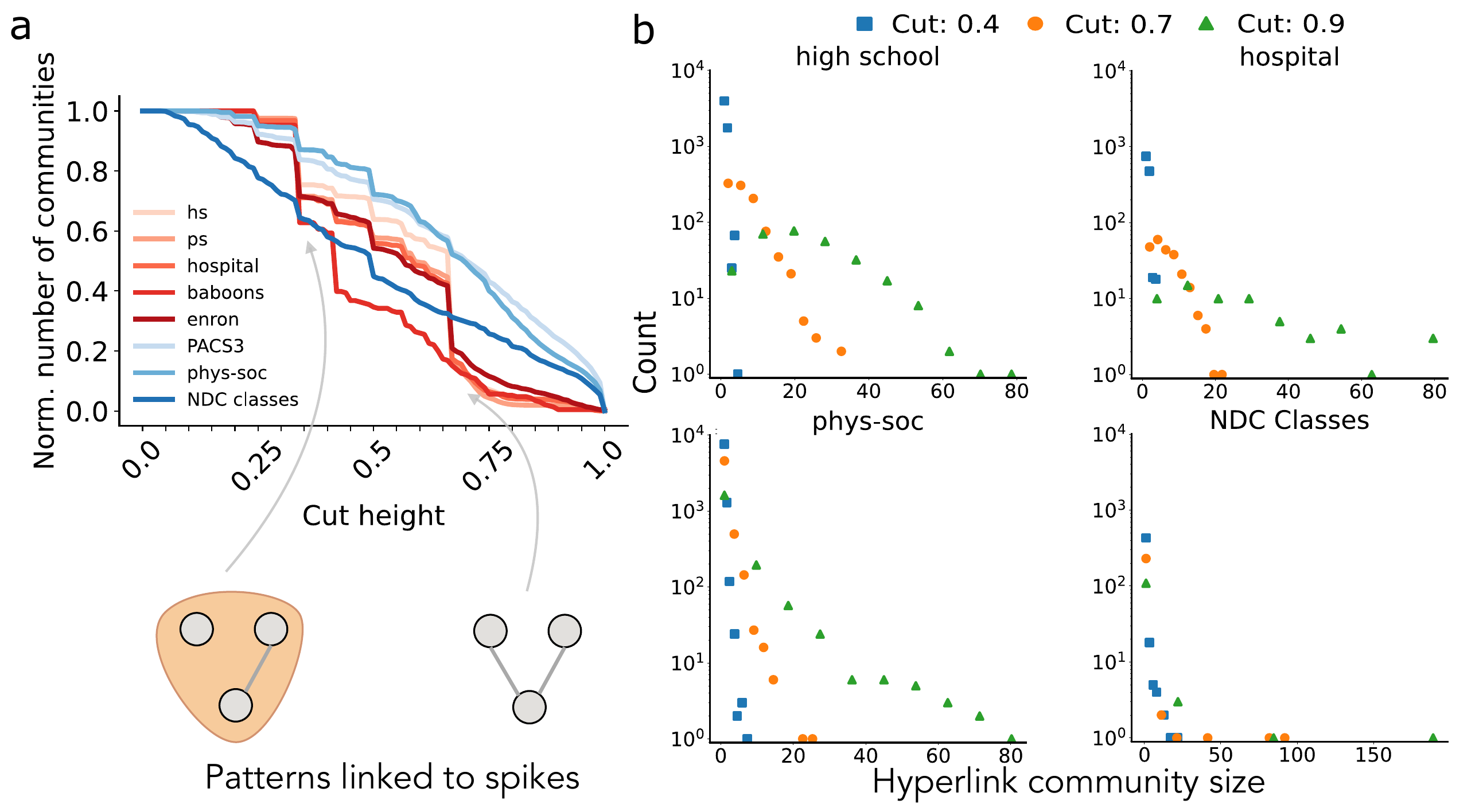}
    \caption{\textbf{Multiscale properties of higher-order networks.} Hierarchical clustering dendrograms can be cut at several thresholds, allowing for the extraction and analysis of hyperlink communities at multiple scales. \textbf{a)} The scaling of the number of hyperlink communities can be interpreted as a fingerprint of the hierarchical organization of group interactions in real-world systems. Due to the over-abundance of certain patterns of overlap between small group interactions social proximity data (red lines) show clear spikes in their curves. \textbf{b)} Evolution of the statistics of the hyperlink communities at different thresholds. Hyperlink community structures can change significantly across scales.}
    \label{fig:fig4}
\end{figure*}

\subsection{Overlapping communities at multiple scales}
\begin{figure*}
    \centering
    \includegraphics[width=\linewidth]{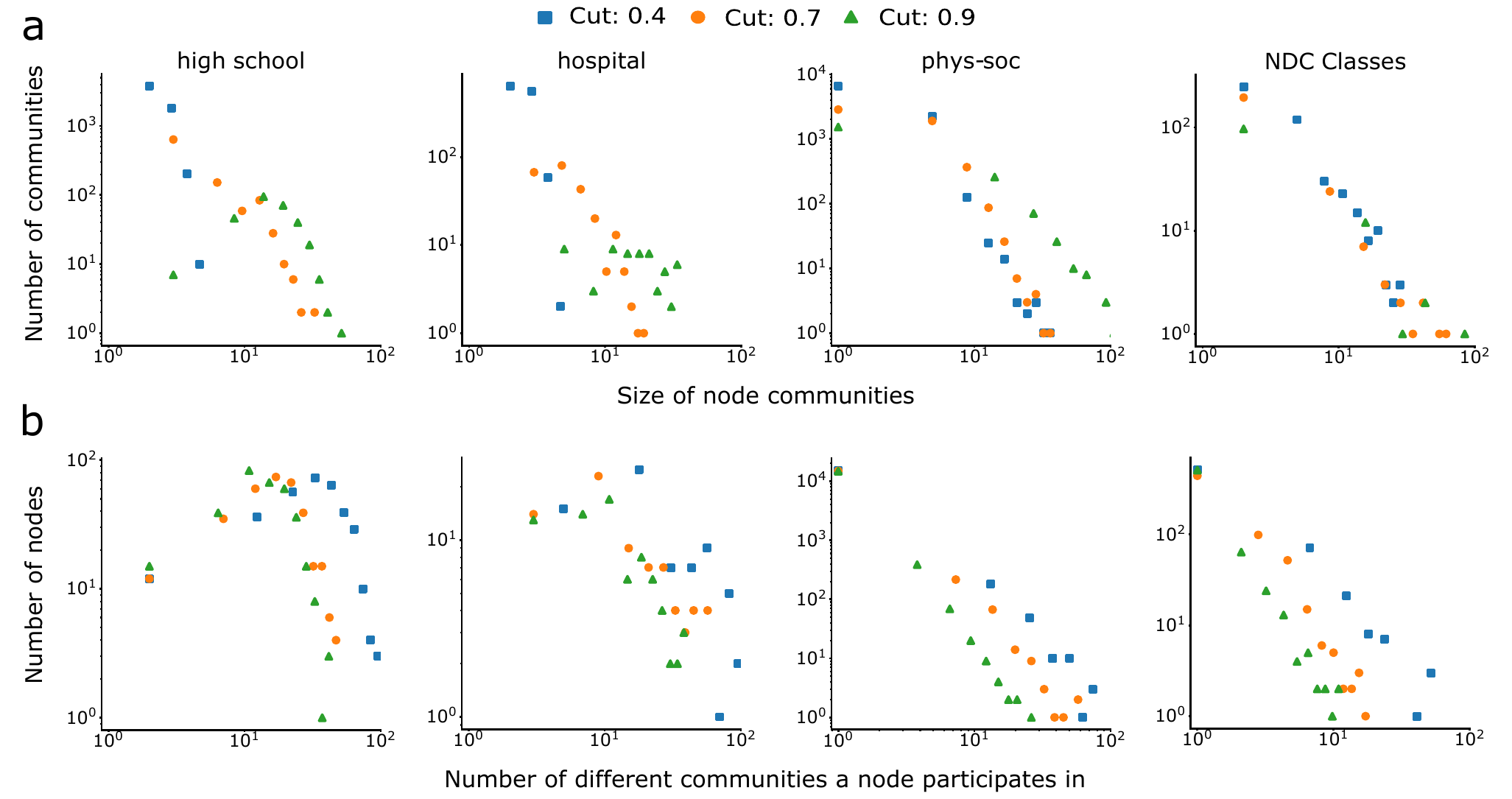}
    \caption{\textbf{Statistics of overlapping communities at multiple scales.} 
 The distribution of node community sizes and node community memberships for several hypergraphs at three different dendrogram thresholds reveals the multiscale overlapping structure of real-world hypergraphs at their mesoscale. The hypergraphs show a wide range of community sizes, generally exponentially distributed, throughout the dendrogram. The distributions of community memberships per node show that nodes tend to participate simultaneously in more communities. This behaviour is consistent across scales. Proximity data has a more pervasive overlapping structure than the other datasets.}
    \label{fig:fig5}
\end{figure*}

Previously we mentioned that by cutting at a certain threshold the dendrogram constructed by applying hierarchical clustering on the hyperlinks distance matrix, not only does the algorithm uniquely assign a community to each hyperlink, but also assigns multiple (possibly overlapping) communities to nodes. By varying the dendrogram-cutting threshold, we can extract overlapping communities across various scales. Having fixed a dendrogram-cutting threshold, for a node $n$, we can define the community membership vector $v_n$ as:

\begin{equation}
    v_n = \{ c(e) : e \in E(n)\}
\end{equation}

where $E(n)$ is the set of hyperlinks in which $n$ participates, and the function $c(e)$ assigns a hyperlink to its community.

In the following, we study the community membership vectors of the nodes across datasets and scales, to quantify node community overlap in real-world hypergraphs. 

In Fig.~5, we analyze the distributions of node community sizes (i.e., the number of nodes participating in a community) and node community memberships (i.e., the communities a node simultaneously participates in) at different cuts of the dendrogram. We find real-world hypergraphs to exhibit a wide spectrum of node community sizes, generally exponentially distributed. The distributions of community memberships per node at different cuts of the dendrogram remain in line across cut heights and show that nodes tend to consistently participate in more than one community at the same time. This suggests that real-world hypergraphs present overlapping structures at multiple scales. Moreover, the different shapes of the distributions highlight once again the different behaviour of the two families of datasets, with proximity data having a more pervasive overlapping structure. 

\begin{figure*}
    \centering
    \includegraphics[width=\linewidth]{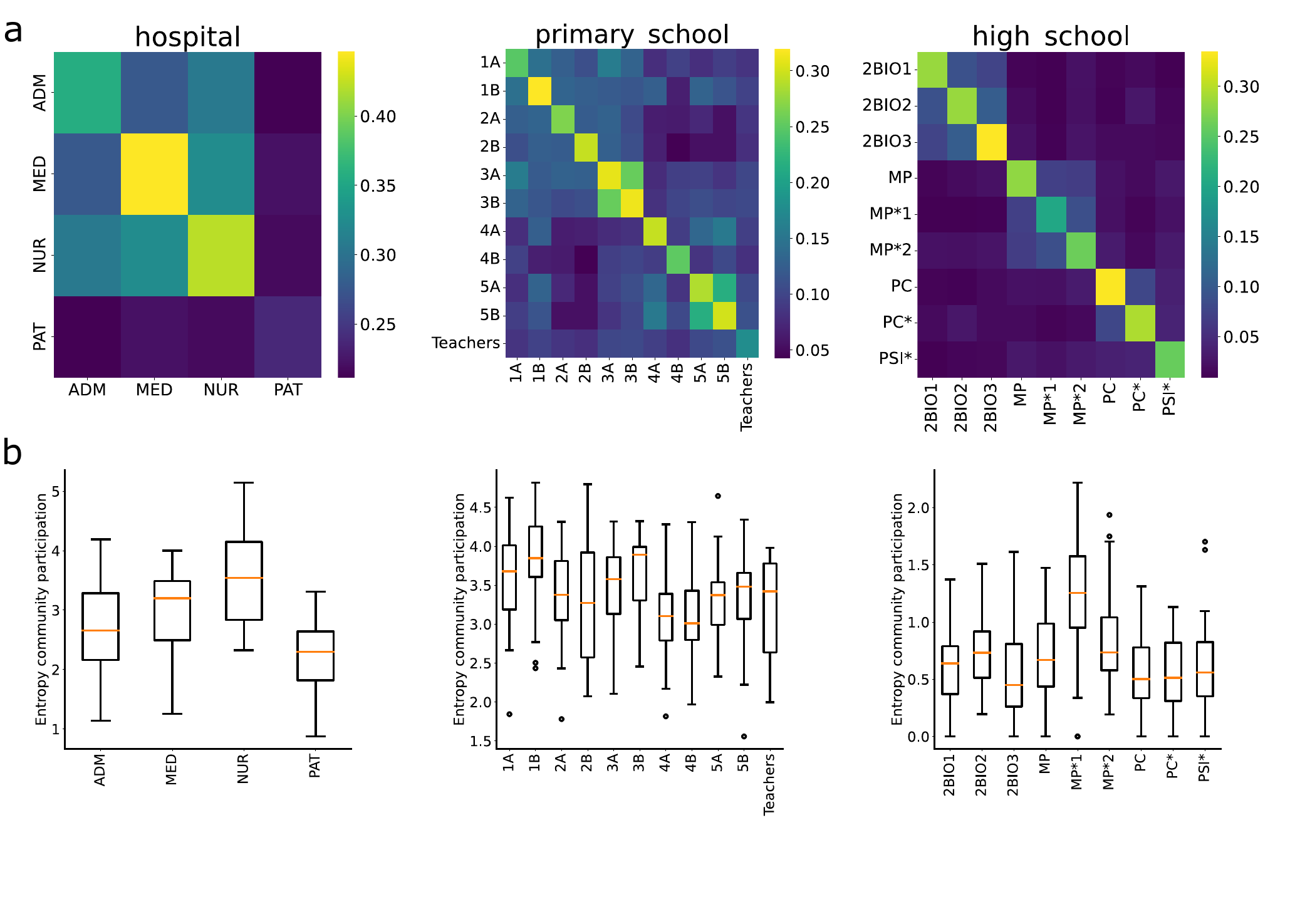}
    \caption{\textbf{Comparing overlapping communities and node metadata.} We select a threshold for cutting the dendrogram and extracting overlapping communities, and compare results with metadata from real-world hypergraphs (role or class). A binary community membership vector is used to identify whether a node participates in a certain community. \textbf{a)} We measure the pairwise similarity between the binary vectors (Jaccard similarity) and build the role-to-role similarity matrices by aggregating similarities of nodes based on their role. Nodes with similar roles or classes tend to share similar community memberships. However, patients in the hospital dataset have low overlapping memberships even with other patients. Moreover, clustering emerges among classes in the primary and high-school datasets, probably because their proximity leads to mixing interactions among different classes. \textbf{b)} We measure the diversity (entropy) of community membership vectors for each role or class averaging nodes with the same role. Certain roles, such as nurses, have a more diverse and pervasive overlap, while patients have less diversified interactions. In school datasets, some classes have more diverse community memberships, possibly due to physical constraints or participation in more activities.}
    \label{fig:fig6}
\end{figure*}

In Fig.~6a, we provide a comparison between the overlapping communities extracted from real-world hypergraphs and the associated metadata. For the hospital dataset, describing proximity interactions among people inside a hospital, each node is characterized by metadata about their role (e.g., nurse). For the high-school and primary-school datasets, describing proximity interactions among students and teachers inside a primary school and a high school, we have information about the class of each individual. For each node $n$, we compute a binary community membership vector. In this vector, for a node $n$, entry $i$ is equal to $1$ if node $n$ participates in the community $i$ at the representative cut, and equal to $0$ otherwise. To identify nodes with similar connectivity patterns, we compute pairwise similarities between binary vectors using the Jaccard similarity. For each dataset, we create the role-to-role similarity matrix $\mathcal{M}$ in which $\mathcal{M}_{ij}$ is the average similarity between the vectors of the units with role $i$ and the units with role $j$. In the hospital dataset, we show that the average similarity of the vectors of nodes with the same role is higher than that of nodes with different roles. This is particularly true for medics and nurses. However, this is not true for specific roles such as patients, who tend not to overlap much neither with other roles or with other patients. In a similar way, we notice that students from the same class in both high school and primary school datasets tend to be more similar than students from different classes. Moreover, some additional form of clustering emerges, probably due to the fact that classes that are more physically close have students that interact more, leading to some communities overlapping among those classes.

In Fig.~6b, we measure the average diversity of the community membership vectors for each role or class. We measure this property by considering the entropy of the community membership vector for each node, averaging nodes with the same role.

Let \( v_i \) be the community membership vector for node \( i \), \( p_{ij} \) be the proportion of memberships of node \( i \) to community \( j \) (i.e., the number of distinct hyperlinks assigned to community $j$ containing node $i$), and \( C \) be the set of communities node \( i \) participates in. The entropy \( H \) for node \( i \) is:

\begin{equation}
H(v_i) = - \sum_{j}^{C} p_{ij} \log(p_{ij})
\end{equation}

Let \( N_r \) be the number of nodes with role \( r \); to average the entropy for nodes of the same role \( r \) we compute:
\begin{equation}
\bar{H}_r = \frac{1}{N_r} \sum_{i \in r} H(v_i)
\end{equation}

We show that, indeed, different roles can have a more diverse set of memberships than others. For example, nurses share a more diverse and pervasive overlap than other roles, while patients are more strict and less overlapping. In the school datasets, there are classes that are more favoured than others in having multiple diverse community memberships, possibly due to physical constraints or participation in more activities. 

\subsection{Cartography of higher-order networks}
\begin{figure*}
    \centering
    \includegraphics[width=\linewidth]{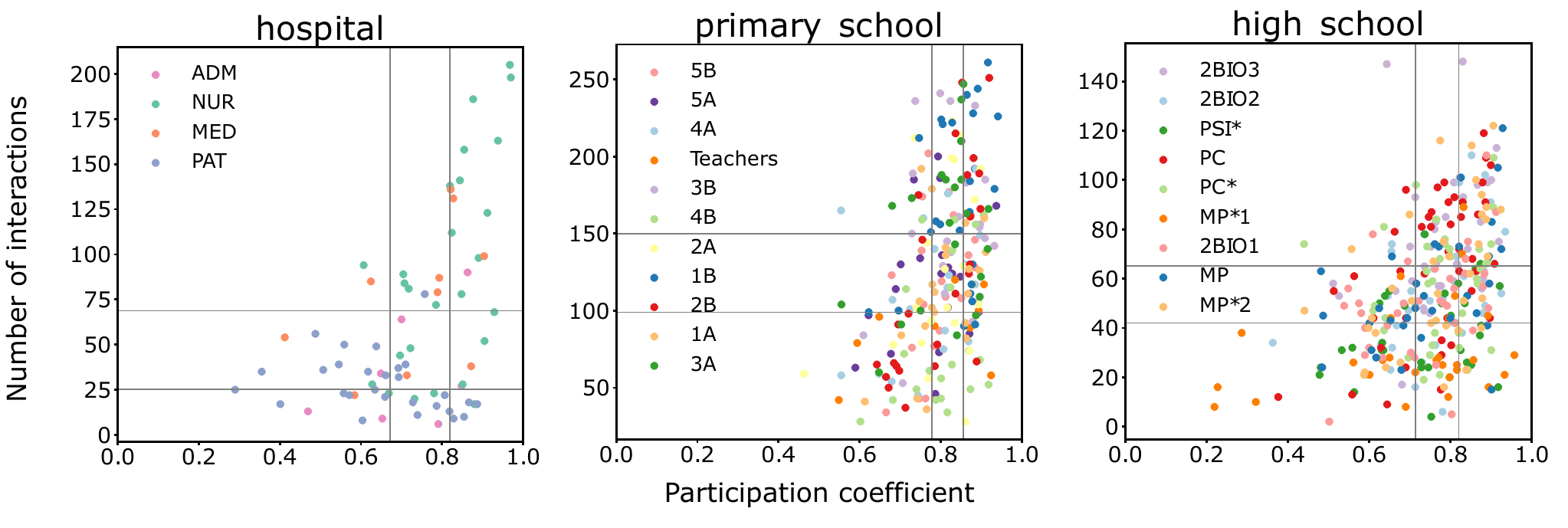}
    \caption{\textbf{Cartography of higher-order networks.} We provide a cartography of higher-order networks, where nodes are classified based both on their number of interactions (hyperdegree) and mixed-membership to hypergraph communities (participation coefficient). This classification yields nine structural roles (hub, non-hub, or peripheral on the y-axis; generalist, non-generalist, or specialist on the x-axis), where complementary information is provided by the two variables. We apply our method to three face-to-face higher-order social systems, showing how it can be used to capture metadata information. As an example, hospital patients tend to be peripherals but range from specialists to generalists. By contrast, in school data, each class has representatives of each structural role.}
    \label{fig:fig7}
\end{figure*}
In the previous sections, we have mentioned that nodes can be characterized in terms of their simultaneous participation in different communities. We have also highlighted that the community memberships of nodes can range from very heterogeneous (i.e., membership vector with high entropy) to very homogeneous (i.e., membership vector with low entropy). Community participation behaviour of nodes can be linked with their hyperdegree to classify nodes into different structural roles. 

We measure the hyperdegree of each node by considering the number of interactions they participate in. Community membership diversity is here computed by considering the participation coefficient of the membership vectors, paying homage to the seminal work by Guimera et al. on the cartography of complex networks~\cite{Guimer2005FunctionalCO} (analogous results are obtained by considering entropy). Such a score of diversity ranges from $0$ (minimum diversity) to $1$ (maximum diversity) and has also been used to characterize heterogeneity in different network structures such as multiplex networks~\cite{battiston2014structural}. In particular, having fixed a scale and computed overlapping communities, the participation coefficient of the membership vector of node $i$ is defined as

\begin{equation}
P_i = \frac{C}{C-1} \left[ \sum_{\alpha=1}^C \left(\frac{k_i^{[\alpha]}}{k_i}\right)^2 \right]
\end{equation}

where $C$ is the number of communities, $k_i$ is the hyperdegree of node $i$ and $k_i^{[\alpha]}$ is the number of interactions of node $i$ that are part of community $\alpha$.

In Fig.~\ref{fig:fig7}, each node is represented as a point in the Cartesian plane with coordinates equal to its hyperdegree and participation coefficient. The nodes are classified into different regions by subdividing both the $x$-axis and the $y$-axis into the $33$rd and $66$th percentiles. On the $y$-axis, a node in the top third is classified as hub, in the middle third as non-hub and in the bottom third as peripheral. On the $x$-axis, a node in the top third is classified as a generalist, in the middle third as a non-generalist and in the bottom third as a specialist. This classification gives a total number of nine structural roles, corresponding to different regions in the plane. Our results highlight that network nodes with different structural roles are indeed relevant for understanding real-world systems. Nodes with a similar hyperdegree can have a very different participation coefficient, and vice versa. Additionally, structural roles can be linked with metadata information. For example, in hospital data, we observe that people with the same jobs or status can have very different behaviours, e.g., patients tend to be peripherals but can range from specialists to generalists. Finally, in school data, in the same class, we have multiple representatives for each structural role. In Appendix~\ref{supp:cartography} we provide more examples of cartography of real-world hypergraph data.

\section{Conclusions}
The analysis of network communities is widely recognized as a fundamental tool for understanding the interplay between the structure and dynamics of complex systems, finding applications in fields such as biology and social network analysis. Network communities are often defined as groups of nodes that are more densely connected to each other within the community than they are to nodes outside of the community. Recently, the framework of hypergraphs has established itself as a fundamental tool to model systems whose interactions are not limited to pairs of nodes but may involve an arbitrary number of them.

Using a dual approach to community detection, in this work we have extended the traditional notion of link communities to hypergraphs, extracting clusters of highly related hyperedges. By defining a measure to determine the distance between two hyperedges and performing hierarchical clustering on top of the pairwise distance matrix of such hyperedges, we studied the dendrogram obtained as the output of such a process on a variety of real-world data. Hyperlink communities naturally highlight the hierarchical and multiscale structure of a higher-order network, at the same time revealing overlaps among node communities. Finally, we have introduced the notion of the cartography of higher-order networks, and classified nodes in different structural roles, i.e., a small number of system-independent roles that depend on the patterns of interactions of the nodes and their scale-specific overlapping community participation. We showed that with this classification we are able to capture information that the hyperdegree alone cannot provide. 
 
Given the interest in link communities in pairwise networks, we believe that hyperlink community detection may serve as a relevant tool for analysing a variety of higher-order data, helping unveil structural patterns which cannot be explored with traditional community detection approaches. In this direction, interesting venues for future research include the evaluation of distance measures for hyperedges alternative to the Jaccard distance used here, and the development of methods to scale the computation of hyperlink distances in very large real-world hypergraphs.

\section*{Acknowledgements} 
F.B. acknowledges support from the Air Force Office of Scientific Research under award number FA8655-22-1-7025. A.M. acknowledges support from the European Union through Horizon Europe CLOUDSTARS project (101086248).

\appendix

\renewcommand{\thefigure}{A\arabic{figure}}
\renewcommand{\thetable}{A\arabic{table}}
\setcounter{figure}{0}
\setcounter{table}{0}

\section{Supplementary Information}
\subsection{Distance measures for hyperedge similarity}
\label{supp:distance}
\begin{figure}[h]
    \centering
    \includegraphics[width=\columnwidth]{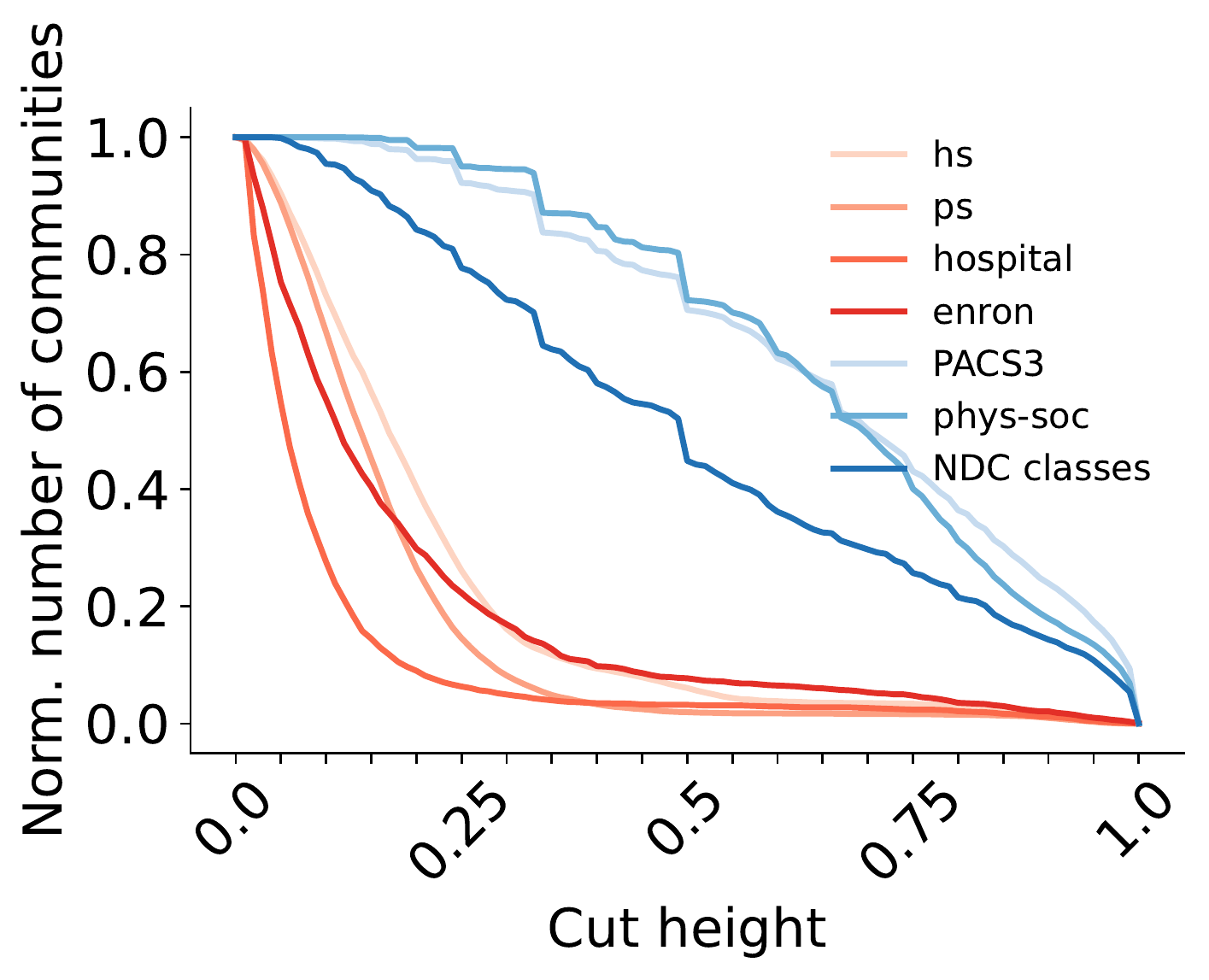}
    \caption{Hierarchical scaling fingerprints computed using a direct generalization of the distance defined in~\cite{ahn2010link}. The separation of two classes of data also holds using this distance.}
    \label{fig:reb-cut}
\end{figure}

In the main text, we use the Jaccard distance because it allows us to investigate patterns of overlapping hyperedges, obtaining an interesting characterization of two different classes of higher-order networks. Moreover, Jaccard distance is coherent with hypergraphs line-graph definitions~\cite{aksoy2020hypernetwork} and the use of graphs line-graph for computing overlapping communities~\cite{evans2009line}. It is also worth noting that our proposed distance is easier to compute than a direct generalization of the distance defined in~\cite{ahn2010link}, although both are computationally expensive. Interestingly, two hyperedges having high/low Jaccard distance tend to have high/low distance as defined in~\cite{ahn2010link}. We measured the correlation (Pearson) between the two distances in the following datasets, selecting a couple of examples from each discovered class: high school ($r=0.96$, $p < 0.01$), Enron mail ($r=0.87$, $p < 0.01$), NDC classes ($r=0.55$, $p < 0.01$) and PACS3 ($r=0.70$, $p < 0.01$). Moreover, Fig.~\ref{fig:reb-cut} shows that results regarding the separation of two classes of higher-order data still hold with a generalization of the distance defined in~\cite{ahn2010link}. The dataset about interactions of baboons is too dense to have a meaningful hierarchical scaling fingerprint using such distance.

\subsection{Datasets}
\label{supp:data}
Summary statistics of the datasets used in our experiments is reported in Table~\ref{tab:datasets}.

\begin{table*}
    \centering
    \begin{adjustbox}{}
        \begin{tabular}{|l|r|r|r|r|r|r|l|}
            \toprule
            Dataset & N & E & $E_2$ & $E_3$ & $E_4$ & $E_5$ & Domain\\
            \midrule
            Phys-soc & 26800 & 15311 & 3700 & 4135 & 2854 & 1613 & Co-auth\\
            NDC\_classes & 1161 & 1088 & 297 & 121 & 125 & 94 & Bio\\
            PACS3 & 33479 & 16977 & 2099 & 2105 & 2590 & 1169 & Co-auth\\
            ENRON & 143 & 1512 & 809 & 317 & 138 & 63 & E-mail\\
            Primary school & 242 & 12704 & 7748 & 4600 & 347 & 9 & Proximity\\
            High school & 327 & 7818 & 5498 & 2091 & 222 & 7 & Proximity\\
            Hospital & 75 & 1825 & 1108 & 657 & 58 & 2 & Proximity\\
            Baboons & 13 & 231 & 78 & 142 & 11 & 0 & Proximity\\
            \bottomrule
        \end{tabular}
    \end{adjustbox}
    \caption{Details of the real-world networked datasets considered for our experiments. Real-world hypergraphs from different domains are described by their number of nodes, total number of hyperedges and number of hyperedges of size $2$, $3$, $4$ and $5$.}
    \label{tab:datasets}
\end{table*}

\subsection{Higher-order cartography for additional datasets}
\label{supp:cartography}
\begin{figure*}
    \centering
    \includegraphics[width=\linewidth]{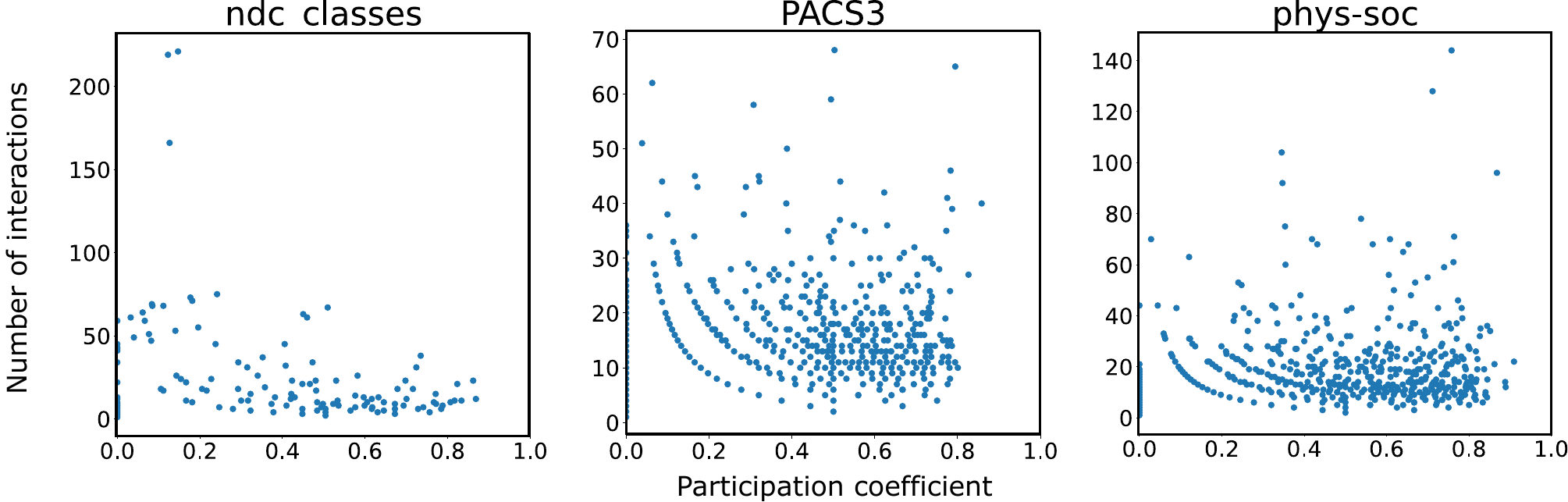}
    \caption{Higher-order cartography for additional datasets.}
    \label{fig:reb-cartography}
\end{figure*}

In Fig.~\ref{fig:reb-cartography} we show the cartography for additional systems. Consistently with the dataset discussed in the main text, the plots show large heterogeneity, with nodes with the same number of interactions spanning very different values of participation coefficients. These relational datasets do not have information about node attributes, and hence -- differently from the social proximity datasets presented in the main texts -- do not allow us to compare higher-order roles for same-class nodes and different-class nodes. In these datasets, the participation coefficient distribution is skewed towards $0$.

\bibliography{biblio}

\end{document}